\documentclass[11pt]{article}

\usepackage{arxiv}

\usepackage[utf8]{inputenc} % allow utf-8 input
\usepackage[T1]{fontenc}    % use 8-bit T1 fonts
\usepackage{hyperref}       % hyperlinks
\usepackage{url}            % simple URL typesetting
\usepackage{booktabs}       % professional-quality tables
\usepackage{amsfonts}       % blackboard math symbols
\usepackage{amsmath}
\usepackage{amssymb}
\usepackage{nicefrac}       % compact symbols for 1/2, etc.
\usepackage{microtype}      % microtypography
\usepackage{lipsum}		% Can be removed after putting your text content
\usepackage{graphicx}
\usepackage{natbib}
\usepackage{doi}

\newcommand{\el}[2]{\varepsilon^{#1}_{#2}}

\title{A Simple Comparison of Biochemical Systems Theory and Metabolic Control Analysis}

\author{ \href{https://orcid.org/0000-0002-3659-6817}{\includegraphics[scale=0.06]{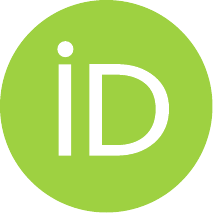}\hspace{1mm}Herbert M.~Sauro}\\
	Department of Bioengineering\\
	University of Washington\\
	Seattle, WA 98195-5061 \\
	\texttt{hsauro@uw.edu} }

\renewcommand{\shorttitle}{BST and MCA}

\hypersetup{
pdftitle={Comparing BST to MCA},
pdfsubject={systems biology},
pdfauthor={Herbert M.~Sauro},
pdfkeywords={Biochemical Systems Theory, Metabolic Control Analysis, Logarithmic gain},
colorlinks,
linkcolor=blue,
citecolor=blue,
urlcolor=blue
}

\begin{document}
\maketitle

\begin{abstract}
This paper explores some basic concepts of Biochemical Systems Theory (BST) and Metabolic Control Analysis (MCA), two frameworks developed to understand the behavior of biochemical networks. Initially introduced by Savageau, BST focuses on system stability and employs power laws in modeling biochemical systems. On the other hand, MCA, pioneered by authors such as Kacser and Burns and Heinrich and Rapoport, emphasizes linearization of the governing equations and describes relationships (known as theorems) between different measures. Despite apparent differences, both frameworks are shown to be equivalent in many respects. Through a simple example of a linear chain, the paper demonstrates how BST and MCA yield identical results when analyzing steady-state behavior and logarithmic gains within biochemical pathways. This comparative analysis highlights the interchangeability of concepts such as kinetic orders, elasticities and other logarithmic gains.
\end{abstract}

% keywords can be removed

\section{Introduction}

In the late 60s and early 1970s Michael Savageau~\citep{savageau1972behavior,savageau1976biochemical} developed an approach called biochemical systems theory (BST). This was a set of mathematical tools to help understand the behavior of biochemical networks. This work arose alongside another similar and independently developed approach in Europe called metabolic control analysis (MCA) by authors Kacser and Burns~\citep{kacser1973control} and Heinrich and Rapoport~\citep{heinrich1974linear}. Both approaches appear at first glance to be different but are actually identical in many respects. Some of the differences arise from different notation but there are also differences in emphasis. BST focuses more on system stability and the use of power laws, while MCA barely mentions stability and is based instead on a direct linearization of the governing equations. Both approaches examine how behavior at the systems level is governed by properties at the local level. MCA also emphasizes theorems that describe relationships between the different measures. Despite their differences, both approaches have provided a wealth of insight into how biochemical networks operate with BST providing a fairly thorough analysis of systems with negative feedback. In this short article, I will show, using some simple examples, the equivalence between the two approaches. No significant new results are presented and the article serves more as an introduction to these two important frameworks.  We will start by considering BST.

A key innovation with BST was the use of power laws as reasonable approximations to use when building models of biochemical networks. For example, given the following single irreversible reaction with substrate $X_1$:
$$ X_1 \stackrel{v_i}{\longrightarrow} $$
the reaction rate, $v_i$, can be written in a power law form:
$$ v_i = \alpha_i X_1^{g_{i1}} $$
where $\alpha_i$ is called the apparent rate constant and the exponent $g_{i1}$ is called the apparent kinetic order. If the kinetic order is one, the equation reduces to the simple first-order mass-action rate law. The power law formulation arises from a Taylor expansion in log space~\citep{savageau1969biochemical} at a specific operating in the kinetic response. The kinetic orders are local sensitivities in log space, that is:
$$ g_{ij} = \frac{\partial v_i}{\partial X_j} \frac{X_j}{v_i} = \frac{\partial \log (v_i)}{\partial \log (X_j)} $$
These are exactly equivalent to the elasticities used in MCA.

BST also uses the symbol $h$ for kinetic orders specifically for degradation steps however this distinction is less important as we'll see when building complete models. Given this formalism is it straightforward to write a set of differential equations for a given model where we assign each reaction a specific power law. For example, consider a four-step linear chain:
$$ X_o \stackrel{v_1}{\longrightarrow} X_1 \stackrel{v_2}{\longrightarrow} X_2 \stackrel{v3}{\longrightarrow} X_3 \stackrel{v_4}{\longrightarrow} $$
We will assume that the species $X_o$ is fixed so that the pathway can sustain a steady-state. The differential equations for this system can be easily written in terms of the production and consumption rates as:
\begin{align}
\begin{split}
\frac{dX_1}{dt} &= v_1 - v_2 \\[4pt]
\frac{dX_2}{dt} &= v_2 - v_3 \\[4pt]
\frac{dX_3}{dt} &= v_3 - v_4
\end{split}
\end{align}
For each $v_i$ we can substitute the appropriate power law to give:
\begin{align}
\begin{split}
\frac{dX_1}{dt} &= \alpha_1 X_o^{g_{10}} - \beta_1 X_1^{h_{11}} \\[4pt]
\frac{dX_2}{dt} &= \beta_1 X_1^{h_{11}} - \beta_2 X_2^{h_{22}} \\[4pt]
\frac{dX_3}{dt} &= \beta_2 X_2^{h_{22}} - \beta_3 X_2^{h_{33}}
\end{split}
\label{eqn:2}
\end{align}
The sign of the kinetic order depends on whether the effector increases (positive) or decreases (negative) a reaction rate. This means that the kinetic orders for inhibitors will be negative in value. In this example they are all positive because increases in substrate concentration will increase the reaction rate.

\begin{figure}[htpb]
\centering
\includegraphics[scale=0.9]{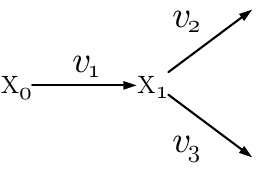}
\caption{Simple branch pathway with two rates $v_2$ and $v_3$ consuming $X_1$.}
\label{fig:branch}
\end{figure}

Note the use of $h$ as the kinetic order for the degradation rate for $v_2$. That is, the symbol $h$ isn't always used to describe degradation steps but, as in this case, it is also used in a consumption step when describing the rate of change of $X_2$. The second equation, $dX_2/dt$, has the same $h$ kinetic order because the same reaction is now a production rate for $X_2$. In fact, in all these equations, the $g$ kinetic orders, other than $g_{10}$ are replaced by $h$ type kinetic orders irrespective of whether we are dealing with a production or consumption step. In general, the distinction between production and consumption kinetic orders does not appear to be so important, especially for reaction networks where each reaction step is explicitly written in the differential equation. Additionally, the notation changes depending on what formalism is being used since BST has a variety of representations it can employ. The two main approaches that are used are the Generalized Mass-Action (GMA) and the S-Systems~\citep{savageau1988introduction} formalism. When dealing with linear chains, as we will do here, the two approaches are identical. They diverge when a model has branches or cycles. In these situations, GMA has explicit terms for each reaction rate. For example, for a simple branch, Figure~\ref{fig:branch}, the rate of change of the branch species will comprise three power laws, one for the production rate and two for the consumption rates emerging from the branch. Equation~\eqref{eqn:branch} shows a GMA representation of the rate of change of the branch species $X_1$. For GMA models, the kinetics orders are generally represented using the $f$ symbol. A more complex example that uses the GMA approach, can be found at~\cite{alves2008mathematical}.
\begin{align}
 \frac{dX_1}{dt} = \alpha_1 X_0^{g_{10}} - \beta_1 X_1^{h_{21}} - \beta_2 X_1^{h_{31}}
 \label{eqn:branch}
\end{align}
For an S-System, which is where most of the utility of BST originates, all power laws associated with the production of a particular species are aggregated. Likewise, all consumption rates are aggregated. This is where $g$ is more likely to be used for the aggregated production rate and $h$ for the aggregated consumption steps to keep track of what is what. Aggregation, however, can be challenging for complex systems that involve branches and cycles. This may involve, for example, computing the weighted kinetic orders~\citep{voit2013biochemical}. For linear chains, the S-system formalism is easy to apply since no aggregation is necessary. This is why GMA and S-Systems are equivalent for linear chains.

\section{Examples}

To illustrate the utility of BST and how it matches MCA, three examples will be considered. The first example is the four-step linear chain of reactions we considered previously. To keep things simple, we will assume that each reaction is irreversible. That is, the product has no influence over its own production. This will mean that each power law only has a single species in the equation which corresponds to the substrate of the reaction. This makes the analysis a little easier. To make a reaction sensitive to product we only need to augment the power law with a product term. For example, to make the first reaction sensitive to product $X_1$, we can add a $X_1$ term with its own kinetic order, $g_{11}$:
\vspace{3pt}
\begin{equation*}
    v_1 = \alpha_1 X_o^{g_{10}} X_1^{g_{11}}
\end{equation*}
We will look at this situation in the next section. Making reactions fully reversible, in the sense that the reaction rate can go negative, is more complex and requires choices to be made when aggregating~\citep{sorribas1989strategies,voit2013biochemical} which we will not consider here.

If we assume the pathway is at steady-state we can set the left-hand side of~\eqref{eqn:2} to zero:
\begin{align}
\begin{split}
0 &= \alpha_1 X_o^{g_{10}} - \beta_1 X_1^{h_{11}} \\[4pt]
0 &= \beta_1 X_1^{h_{11}} - \beta_2 X_2^{h_{22}} \\[4pt]
0 &= \beta_2 X_2^{h_{22}} - \beta_3 X_2^{h_{33}}
\end{split}
\end{align}
Rearrangement and taking logarithms on both sides~\citep{savageau1969biochemical,savageau1971concepts} results in a set of linear equations that can be solved for $\log X_i$. If we designate $\log X_i$ with the symbol $y_i$, these solutions can be written as follows:
\begin{align}
\begin{split}
y_1 &= (g_{10} y_0 - b_1)/h_{11} \\[3pt]
y_2 &= (g_{10} y_0 - b_1 - b_2)/h_{22} \\[3pt]
y_3 &= (g_{10} y_0 - b_1 - b_2 - b_3)/h_{22}
\label{eqn:4}
\end{split}
\end{align}
The $b_i$ terms are the log of the ratio of $\beta_i$ to $\alpha_i$ and $y_i$ as indicated before are the log of the $X_i$ variables. We note again that in this example it is assumed that each reaction is not influenced by its product. The log transformation of the power law representation is a key innovation in BST and allows one to derive analytically the steady-state solution from the original differential equations. This formulation is unique to BST.

The fact that the $y_i$ terms are the log of the $X_i$, makes it very easy to compute the logarithmic gains by simple differentiation of these expressions. Both BST and MCA focus much of their attention on deriving logarithmic gains and it is worth briefly saying something about such quantities. Interestingly the term `logarithmic gain' is not widely used in the literature even though it expresses accurately the concept. The earliest reference I could find to this term is from one of Savageau's earliest papers~\citep{savageau1971concepts}. Similar concepts to the logarithmic gain appear in economics in the form of supply and demand elasticities which is also the origin of the term `elasticity' in MCA. A logarithmic gain is in simple terms a ratio of relative changes. In differential calculus, we are familiar with the ratio of absolute changes, which in the limit become differentials. In a similar fashion logarithmic gains are defined as the limit of infinitesimal relative changes. Experimentally, logarithmic gains are much more useful because they are unit-less and experiments in cell and molecular biology are often expressed as ratios of unit-less fold changes. Given a variable $Y$ and input variable $X$. the logarithmic gain of how $X$ can influence $Y$ is described as a logarithmic gain in a variety of forms:
$$ \text{logarithmic gain } = \frac{dY}{dX} \frac{X}{Y} = \frac{d\log Y}{d \log X} \approx \frac{Y \%}{X \%} $$
In BST, the logarithmic gain is also symbolized using $L_{YX}$ and in MCA as $L^Y_X$ where $L$ is a different letter depending on the gain under consideration. Numbers may also be used instead of letters for both BST and MCA. BST also uses an additional gain called the sensitivity. This measures how a given system property such as a concentration is affected by a change in a system parameter such as a kinetic order. This is symbolized using $S_{Lh}$. There is not standardized notation but generally, the aforementioned notation is commonly found in the literature. 

In log form, it is easy to derive logarithmic gain if the equation under study is already in log form. This is the case with the transformation done by BST.

For example, if we wish to investigate how $X_o$ influences the steady-state levels of $X_i$, we only have to differentiate each $y_i$ with respect to $y_o$, because by design $y_o = \log X_o$ and $y_i = \log X_i$. That is:
\begin{align}
\frac{\partial y_1}{\partial y_0} = \frac{\partial \log X_1}{\partial \log X_o}  
\end{align}
If we differentiate $y_1 = (g_{10} y_0 - b_1)/h_{11}$ with respect to $y_0$, we obtain (noting that the $b$ terms are constant):
\begin{align}
\frac{\partial \log X_1}{\partial \log X_o}  = \frac{g_{10}}{h_{11}} 
\end{align}
As mentioned before logarithmic gains are useful because they can be interpreted as ratios of percentage changes which makes them experimentally measurable. The other two logarithmic gains can be derived in the same way to give the complete set of relations between $X_o$ and the three intermediates:
\begin{align}
\begin{split}
\frac{\partial \log X_1}{\partial \log X_o} &= \frac{g_{10}}{h_{11}} \\[4pt]
\frac{\partial \log X_2}{\partial \log X_o} &= \frac{g_{10}}{h_{22}} \\[4pt]
\frac{\partial \log X_3}{\partial \log X_o} &= \frac{g_{10}}{h_{33}}
\label{eqn:789}
\end{split}
\end{align}
From these equations, we see a clear pattern that makes them easy to extend to any length pathway. These results are well known~\citep{savageau1972behavior} and illustrate one of the key features of BST which is the ability to find analytical solutions for the steady-state using simple linear algebra from which the logarithmic gains are easily derived. This approach works extremely well for linear chains and linear chains with any manner of feedback loops. 

Because the kinetic orders are positive values, each of the logarithmic gains is also positive. This simply means that increasing $X_0$ will increase the steady-state concentrations of $X_1, X_2$, and $X_3$ which makes logical sense.

\subsection*{Comparison with MCA}

We now compare the BST derivation with that from MCA. The first difference is that in MCA~\citep{kacser1995control,heinrich1974linear,sauroMCABook} the kinetic orders are called the elasticity coefficients and like BST, logarithmic gains can be derived in terms of the elasticities. 

Over the years a variety of ways have been developed to derive the logarithmic gains~\citep{sauroMCABook}. One of these approaches is given in the appendix which uses implicit differentiation of the differential equations when written in functional form. The solutions using this method are given below:
\begin{align}
R^{X_1}_{X_0} = \frac{\el{1}{0}}{\el{2}{1}}, \qquad R^{X_2}_{X_0} = \frac{\el{1}{0}}{\el{3}{2}}, \qquad R^{X_3}_{X_0} = \frac{\el{1}{0}}{\el{4}{3}} 
\label{eqn:9}
\end{align}
We can compare these equations with the ones derived by BST~\eqref{eqn:789}. It is clear there is a direct equivalence. The kinetic order $g_{10}$ is equivalent to the elasticity $\el{1}{0}$ and the kinetic orders $h_{11}, h_{22}$, and $h_{33}$ are equivalent to the elasticities $\el{2}{1}, \el{3}{2}$, and $\el{4}{3}$ respectively.

MCA categories the logarithmic gains depending on their role and also introduces additional logarithmic gains not present in BST although in principle could be computed by BST. BST does have the additional sensitivity measure that was mentioned above. Two of these logarithmic gains are relevant here, they are the concentration control coefficients and concentration response coefficients.  The response coefficient describes how an external factor such as $X_0$ influences the steady-state. The response coefficient is equivalent to the logarithmic gain derived above using BST. In MCA terminology the response coefficient of $X_1$ with respect to $X_0$ is given by $R^{X_1}_{X_0}$:
\begin{align}
R^{X_1}_{X_0} = \frac{\partial \log X_1}{\partial \log X_0} = L_{X_1,X_0} = L_{ik}
\end{align}
As noted before, un BST the logarithmic gains are also given the symbol $L_{ik}$ which describes how variable $X_i$ is influenced by external factor $k$. 

{\bf In terms of the response coefficient} A theorem in MCA states that a given response coefficient is related to the product of an elasticity and a control coefficient as follows:
\begin{align}
R^{X_i}_{X_0} = C^{X_i}_{v_i} \el{v_i}{X_o}
\end{align}
where $C^{X_i}_{v_i}$ is the control coefficient of $X_i$ with respect to reaction $v_i$. A concentration control coefficient describes how a perturbation in a reaction (for example via a change in enzyme activity) can lead to a change in a steady-state concentration. Thus $C^{X_j}_{v_i}$ describes how reaction $v_i$ influences species $X_j$.

In MCA the control coefficient for the first step, $v_1$, with respect to $X_1$ can be shown to equal:
\begin{align}
C^{X_1}_{v_1} = \frac{1}{\el{2}{1}}
\end{align}
This allows us to write the response coefficient as:
\begin{align}
R^{X_1}_{X_0} =  \frac{\el{1}{X_o}}{\el{2}{1}}
\end{align}
which is again equivalent to the same equation from BST. The equivalence between all the equations is summarized below:
\begin{align}
L_{X_1,X_0} &= \frac{g_{10}}{h_{11}} = R^{X_1}_{X_0} = \frac{\el{1}{X_o}}{\el{2}{1}} \\[4pt]
L_{X_2,X_0} &= \frac{g_{10}}{h_{22}} = R^{X_2}_{X_0} = \frac{\el{1}{X_o}}{\el{3}{2}} \\[4pt]
L_{X_3,X_0} &= \frac{g_{10}}{h_{33}} = R^{X_3}_{X_0} = \frac{\el{1}{X_o}}{\el{4}{3}}
\end{align}

\subsection*{Including Product Inhibition}

The previous example assumed that each reaction was unaffected by its product. In reality, this is not necessarily the case~\citep{shen2020combined}. In this example, we will include product inhibition. To keep things simple, let's look at a two-step pathway but it can be extended to any length pathway:
$$ X_o \stackrel{v_1}{\longrightarrow} X_1 \stackrel{v_2}{\longrightarrow} $$
If $X_1$ can product inhibit $v_1$, the power law for $v_1$ must be modified to:
$$ v_1 = \alpha_1 X_0^{g_{10}} X_1^{h_{01}} $$
The differential equation for $X_1$ then becomes;
$$ \frac{dX_1}{dt} = \alpha_1 X_0^{g_{10}} X_1^{h_{01}} - \beta_1 X_1^{h_{11}} $$
Setting this equation to zero, taking logarithms on both sides and rearranging to solve for $y_1$ as before yields:
$$ y_1 = \frac{g_{10} y_0 - b_1}{h_{11} - h_{01}} $$
Compared to equation~\eqref{eqn:4} we see there is an extra term in the denominator. As before we can derive the logarithmic gain by differentiation to obtain:
$$ \frac{\partial \log X_1}{\partial \log X_o}  = \frac{g_{10}}{h_{11} - h_{01}} $$
If we use MCA, we obtain the same relationship:
$$ R^{X_1}_{X_0} = \frac{\el{1}{0}}{\el{2}{1} - \el{1}{1}} $$
with the kinetic orders replaced by elasticities. Both BST and MCA produce the same results.

\subsection*{Including Negative Feedback}

\begin{figure}
\centering
\includegraphics[scale=0.65]{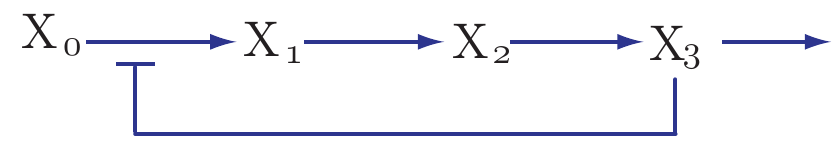}
\caption{Four-step pathway with negative feedback.}
\label{fig:feedback}
\end{figure}

One final example worth looking at is when there is a negative feedback loop. Here we will look at the results common to both as the BST community has done a more extensive analysis of negative feedback. For comparison purposes, we will use the same feedback model found in~\citep{savageau1972behavior}. This model is a four-step linear chain with a feedback loop from $X_3$ to the first step (Figure~\ref{fig:feedback}).

Following~\citep{savageau1972behavior} the power law for the inhibited step is given by:
$$ v_1 = \alpha_1 X_0^{g_{10}} X_3^{g_{13}} $$
This is very similar to the way product inhibition was dealt with in the previous section. Using the same approach as before, we write out the differential equations, setting them to zero, and taking the log on both sides yields:
\begin{align}
b_1 &=g_{10} y_0-h_{11} y_1+g_{13} y_3 \\
b_2 &=h_{11} y_1-h_{22} y_2 \\
b_3 &=h_{22} y_2-h_{33} y_3
\end{align}
Solving for the $y$ terms gives:

\begin{align}
y_1 &= \frac{b_1\ h_{33}+ b_2\ g_{13} + b_3\ g_{13} - g_{10}\ h_{33}\ y_o}{h_{11} (g_{13} -  h_{33})} \\[5pt]
y_2 &= -\frac{h_{33} (b_1 + b_2 - g_{10}\ y_o) + b_3\ g_{13}}{h_{22}\ (g_{13} - h_{33})} \\[5pt]
y_3 &= \frac{b_1 + b_2 + b_3 - g_{10}\ y_o}{g_{13} - h_{33}}
\end{align}
In case you're wondering, there is no $y_o$ solution because $y_o$ is the input species which is fixed. To get the logarithmic gains we differentiate these with respect to $y_o$ to yield~\citep{savageau1972behavior}:
\begin{align}
\frac{\partial y_1}{\partial y_0} &= \frac{\partial \log X_1}{\partial \log X_o} = \frac{h_{33}}{\left(h_{33}-g_{13}\right)} \frac{g_{10}}{h_{11}} \\[5pt]
\frac{\partial y_2}{\partial y_0} &= \frac{\partial \log X_2}{\partial \log X_o} = \frac{h_{33}}{\left(h_{33}-g_{13}\right)} \frac{g_{10}}{h_{22}} \\[5pt]
\frac{\partial y_3}{\partial y_0} &= \frac{\partial \log X_3}{\partial \log X_o} = \frac{h_{33}}{\left(h_{33}-g_{13}\right)} \frac{g_{10}}{h_{33}}
\end{align}
The same analysis (see appendix) can be done using MCA to give:
\begin{align}
\begin{split}
R^{X_1}_{X_0} &=  \frac{\varepsilon^{4}_{3}}{(\varepsilon^{4}_{3} - \varepsilon^{1}_{3})} \frac{\varepsilon^{1}_{0}}{\varepsilon^{2}_{1}} \\[5pt]
R^{X_2}_{X_0} &= \frac{\varepsilon^{4}_{3}}{(\varepsilon^{4}_{3} - \varepsilon^{1}_{3})}  \frac{\varepsilon^{1}_{0}}{\varepsilon^{3}_{2}} \\[5pt]
R^{X_3}_{X_0} &= \frac{\varepsilon^{4}_{3}}{(\varepsilon^{4}_{3} - \varepsilon^{1}_{3})} \frac{\varepsilon^{1}_{0}}{\varepsilon^{4}_{3}}
\label{eqn:26}
\end{split}
\end{align}

As we can see these equations are exactly the same as those derived via BST.

%$$ d = \varepsilon^{2}_{1} \varepsilon^{3}_{2} \varepsilon^{4}_{3} - \varepsilon^{1}_{3} \varepsilon^{2}_{1} \varepsilon^{3}_{2} $$

%$$ -\varepsilon^{1}_{1} \varepsilon^{2}_{2} \varepsilon^{3}_{3} + \varepsilon^{1}_{1} \varepsilon^{2}_{2} \varepsilon^{4}_{3} - \varepsilon^{1}_{1} \varepsilon^{3}_{2} \varepsilon^{4}_{3} - \varepsilon^{1}_{3} \varepsilon^{2}_{1} \varepsilon^{3}_{2} + \varepsilon^{2}_{1} \varepsilon^{3}_{2} \varepsilon^{4}_{3}$$

\section{Discussion}

This article discusses the development and comparison of two frameworks for understanding biochemical networks: Biochemical Systems Theory (BST) and Metabolic Control Analysis (MCA). BST, introduced by Savageau, focuses on developing mathematical tools to analyze biochemical network behavior, particularly emphasizing system stability and the use of power laws. MCA, pioneered by Kacser and Burns, and Heinrich and Rapoport, approaches the same problem but with a different emphasis, based on linearization of governing equations and examining relationships between the various logarithmic gains.

There is a strong equivalence between BST and MCA despite their apparent differences. This is shown through three simple examples of a linear chain of biochemical reactions. Both frameworks offer similar is not identical insights into how biochemical networks operate.

The key innovation of BST lies in its use of power laws to model biochemical reactions, allowing for the straightforward formulation into systems of differential equations. These equations facilitate the analysis of steady-state concentrations and logarithmic gains, which represent how changes in a given parameter affect variables such as concentrations in the system. 

Comparing the derivations of logarithmic gains in BST and MCA, it becomes evident that they yield equivalent results, albeit expressed differently. Both frameworks provide a way to quantify the influence of external factors on steady-state concentrations and highlight how local properties contribute to system-wide behavior. Due to the way BST uses aggregation for more complex networks, I feel working with more complex networks is better handled by MCA because the entire procedure can be automated by computer~\citep{christensen2018pyscestoolbox} directly from an SBML model~\citep{hucka2003systems}.

\section*{Acknowledgements}
This work was partly supported by DOE award DESC0023091.

\section*{Note to readers} I welcome critiques of this paper and can update the text accordingly. 

\bibliographystyle{apsr}
\bibliography{references}

\section*{Appendix}

One way to derive the control coefficients in MCA is to use implicit differentiation of the steady-state equations. We start by writing the differential equations in functional form:
\begin{align}
\frac{dX_1}{dt} &= v_1 (X_0) - v_2 (X_1 (X_0)) \\[4pt]
\frac{dX_2}{dt} &= v_2 (X_1 (X_0)) - v_3 (X_2 (X_0)) \\[4pt]
\frac{dX_3}{dt} &= v_3 (X_3 (X_0)) - v_4 (X_4 (X_0)) 
\end{align}
For example, $v_2 (X_1 (X_0))$ means that the rate of reaction $v_2$ is a function of substrate $X_1$. However, $X_1$ in turn is a function of the fixed input substrate $(X_0)$.

We set the equations to zero to indicate we are interested in the steady-state:
\begin{align}
0 &= v_1 (X_0) - v_2 (X_1 (X_0)) \\[4pt]
0 &= v_2 (X_1 (X_0)) - v_3 (X_2 (X_0)) \\[4pt]
0 &= v_3 (X_2 (X_0)) - v_4 (X_3 (X_0)) 
\end{align}
Since we are concerned with how $X_0$ influences the steady-state concentrations, every steady-state species must be a function of $X_0$. To obtain the logarithmic gains we differentiate each equation with respect to $X_0$:
\begin{align}
0 &= \frac{\partial v_1}{\partial X_0} - \frac{\partial v_2}{\partial X_1} \frac{dX_1}{dX_0} \\[4pt]
0 &= \frac{\partial v_2}{\partial X_1} \frac{dX_1}{dX_0} - \frac{\partial v_3}{\partial X_2} \frac{dX_2}{dX_0} \\[4pt]
0 &= \frac{\partial v_3}{\partial X_2} \frac{dX_2}{dX_0} - \frac{\partial v_4}{\partial X_3} \frac{dX_3}{dX_0} 
\end{align}
These equations form a set of simultaneous equations that can be solved for the three derivatives $dX_1/dX_0, dX_2/dX_0$, and $dX_3/dX_0$. For example, the first equation can be solved for $dX_1/dX_0$:
$$ \frac{dX_1}{dX_0} = \frac{\partial v_1}{\partial X_o}\bigg/\frac{\partial v_2}{\partial X_1} $$
The logarithmic gain can be obtained by scaling both sides, noting that at steady-state $v_1 = v_2$:
\begin{align}
    \frac{dX_1}{dX_0} \frac{X_0}{X_1} = 
    \frac{\partial v_1}{\partial X_0} \frac{X_0}{v_1} \bigg/
    \frac{\partial v_2}{\partial X_1}  \frac{X_1}{v_2}
\end{align}
The terms on the right-hand side are the elasticities, $\el{1}{0}$ and $\el{2}{1}$ respectively so that the equation can be written as:
\begin{align}
\frac{dX_1}{dX_0} \frac{X_0}{X_1} = \frac{\el{1}{0}}{\el{2}{1}} 
\end{align}
In MCA the left-hand term is called the response coefficient and designated $R^{X_1}_{X_0}$. The remaining unknowns can be solved in the same way leading to the equation shown in the main text~\eqref{eqn:9}.

{\bf\large Negative Feedback}

Adding negative feedback adds a term to $v_1$ due to the negative feedback via $X_3$. Compared to a system without negative feedback, only the first differential equation is modified as shown below:

\begin{align}
\frac{dX_1}{dt} &= v_1 (X_0, X_3 (X_0)) - v_2 (X_1 (X_0)) \\[4pt]
\frac{dX_2}{dt} &= v_2 (X_1 (X_0)) - v_3 (X_2 (X_0)) \\[4pt]
\frac{dX_3}{dt} &= v_3 (X_3 (X_0)) - v_4 (X_4 (X_0)) 
\end{align}

As before we set the equations to zero to find the steady-state:

\begin{align}
0 &= v_1 (X_0, X_3 (X_0)) - v_2 (X_1 (X_0)) \\[4pt]
0 &= v_2 (X_1 (X_0)) - v_3 (X_2 (X_0)) \\[4pt]
0 &= v_3 (X_3 (X_0)) - v_4 (X_4 (X_0)) 
\end{align}

We implicitly differentiate with respect to $X_0$ to obtain: 

\begin{align}
0 &= \frac{\partial v_1}{\partial X_0} + \frac{\partial v_1}{\partial X_3} \frac{dX_3}{dX_0} - \frac{\partial v_2}{\partial X_1} \frac{dX_1}{dX_0} \\[4pt]
0 &= \frac{\partial v_2}{\partial X_1} \frac{dX_1}{dX_0} - \frac{\partial v_3}{\partial X_2} \frac{dX_2}{dX_0} \\[4pt]
0 &= \frac{\partial v_3}{\partial X_2} \frac{dX_2}{dX_0} - \frac{\partial v_4}{\partial X_3} \frac{dX_3}{dX_0} 
\end{align}

This forms a set of linear equations that can be solved for $dX_1/dX_0, dX_2/dX_0$, and $dX_3/dX_0$ to give:

\begin{align}
\frac{dX_1}{dX_0} &= \dfrac{\dfrac{\partial v_1}{\partial X_0} \dfrac{\partial v_4}{\partial X_3} }{\left(\dfrac{\partial v_4}{\partial X_3} - \dfrac{\partial v_1}{\partial X_3}\right) \dfrac{\partial v_2}{\partial X_1} } \\[5pt]
\frac{dX_2}{dX_0} &= \dfrac{\dfrac{\partial v_1}{\partial X_0} \dfrac{\partial v_4}{\partial X_3} }{\left(\dfrac{\partial v_4}{\partial X_3} - \dfrac{\partial v_1}{\partial X_3}\right) \dfrac{\partial v_3}{\partial X_2} } \\[5pt]
\frac{dX_3}{dX_0} &= \dfrac{1}{\left(\dfrac{\partial v_4}{\partial X_3} - \dfrac{\partial v_1}{\partial X_3}\right) } 
\end{align}

These are unscaled expressions but it is straightforward to convert these to logarithmic gains by multiplying both sides by $X_0$ and dividing both sides by either $X_1, X_2$ or $X_3$ depending on the equation. Because the pathway is linear, at steady-state all rates are equal, thus $v_1 = v_2 = v_3 = v_4$ allowing us to scale the elasticities by $v$. The result is the following scaled expressions as given in the main text~\eqref{eqn:26}:

$$ R^{X_1}_{X_0} =  \frac{\varepsilon^{4}_{3}}{(\varepsilon^{4}_{3} - \varepsilon^{1}_{3})} \frac{\varepsilon^{1}_{0}}{\varepsilon^{2}_{1}} $$

$$ R^{X_2}_{X_0} = \frac{\varepsilon^{4}_{3}}{(\varepsilon^{4}_{3} - \varepsilon^{1}_{3})}  \frac{\varepsilon^{1}_{0}}{\varepsilon^{3}_{2}}  $$

$$ R^{X_3}_{X_0} = \frac{\varepsilon^{4}_{3}}{(\varepsilon^{4}_{3} - \varepsilon^{1}_{3})} \frac{\varepsilon^{1}_{0}}{\varepsilon^{4}_{3}} $$

\end{document}